\pdfoutput=1
\documentclass[aps,pre,twocolumn, amsfonts,showpacs,superscriptaddress]{revtex4-1}
\usepackage{mathptmx}
\usepackage{epsfig,amsopn}
\usepackage{graphicx}
\usepackage{amsmath,amssymb}
\usepackage{natbib}
\usepackage{braket}
\usepackage{bm}

\newcommand{\eref}[1]{Eq.~(\ref{#1})}%
\newcommand{\fref}[1]{Fig.~\ref{#1}} %
\newcommand{\Fref}[1]{Figure~\ref{#1}}%

\begin{document}

\title{Dynamical transition in the temporal relaxation of stochastic
processes under resetting}

\author{Satya N. Majumdar}
\affiliation{Univ. Paris-Sud, CNRS, LPTMS, UMR 8626, Orsay F-01405, France}
\author{Sanjib Sabhapandit}
\affiliation{Raman Research Institute, Bangalore 560080, India}
\author{Gr\'egory Schehr}
\affiliation{Univ. Paris-Sud, CNRS, LPTMS, UMR 8626, Orsay F-01405, France}

\date{\today}

\begin{abstract}

A stochastic process, when subject to resetting to its initial
condition at a constant rate, generically reaches a non-equilibrium
steady state. We study analytically how the steady state is approached
in time and find an unusual relaxation mechanism in these systems. We
show that as time progresses, an inner core region around the
resetting point reaches the steady state, while the region outside the
core is still transient. The boundaries of the core region grow with
time as power laws at late times with new exponents.  Alternatively,
at a fixed spatial point, the system undergoes a dynamical transition
from the transient to the steady state at a characteristic space
dependent timescale $t^*(x)$.  We calculate analytically in several
examples the large deviation function associated with this
spatio-temporal fluctuation and show that generically it has a second
order discontinuity at a pair of critical points characterizing the
edges of the inner core. These singularities act as separatrices
between typical and atypical trajectories. Our results are verified in
the numerical simulations of several models, such as simple diffusion
and fluctuating one-dimensional interfaces.

\end{abstract}


\maketitle

{\em Introduction.}---
Consider a stochastic process evolving under some given dynamics that
does not lead to a time-independent stationary state. The prototypical
example being the position of a diffusive particle---which has a
Gaussian distribution with its width growing as the square root of the
time, inferring the absence of a steady state. Now, imagine that by
another mechanism, the dynamics is repeatedly being interrupted and
recommenced at random times from the initial condition. A general
interesting question is: how does such a stochastic resetting affect
the temporal evolution of the system?

Examples of stochastic resetting are found in a wide variety of
situations. For example, while looking for a friend in a crowded
tourist place, after an unsuccessful search over some time period, one
often returns to the most favorite hangout and restarts the search
process again. In the ecological context, the animal movements are
often modeled by stochastic
processes~\cite{turchin,Vishwanathan:book}.  The movements of foraging
animals usually involve local diffusive search for food, interrupted
by long range non-local resetting moves to relocate to other areas as
well as to return to their nests, followed by restart of the search
process~\cite{benichou}. For instance, the mobility data of
free-ranging capuchin monkeys is described quite well by a model of
random walks with preferential relocations to places visited in the
past~\cite{Boyer:14}.  Similar notions can be also found in biological
contexts, where organisms use stochastic resetting or switching
between different phenotypic states to adapt to fluctuating
environments~\cite{Kussell:05, Kussell:05b, Reingruber:09, Visco:10,
Reingruber:11}.  In computer science, random walks with stochastic
restarts turns out to be a useful strategy to optimize search
algorithms in hard combinatorial problems~\cite{Lovasz:96,
Montanari:02, Konstas:09}.

The stochastic resetting mechanism has been shown to have rather rich
and dramatic effects on the diffusion
process~\cite{Evans:11,Evans:11b,Evans:13,Whitehouse:13,Evans:14,Arnab:14},
as well as on long range jump processes such as L\'evy
flights~\cite{Kusmierz14}. While in the absence of the resetting, the
diffusion in free space does not have a stationary state, a non-zero
rate of resetting to a fixed position leads to a non-equilibrium
stationary state (NESS) with non-Gaussian fluctuations, in the time
$t\to \infty$ limit.  Similarly, an extended system like a fluctuating
interface, evolving under its own dynamics and is reset at a constant
rate to its initial configuration, approaches at late times a NESS
with a nontrivial interfacial height
distribution~\cite{Gupta:14}. Resetting induced NESS has also been
studied in other many body systems such as coagulation-diffusion
processes~\cite{Durang:14}.

\begin{figure}
\includegraphics[width=\hsize]{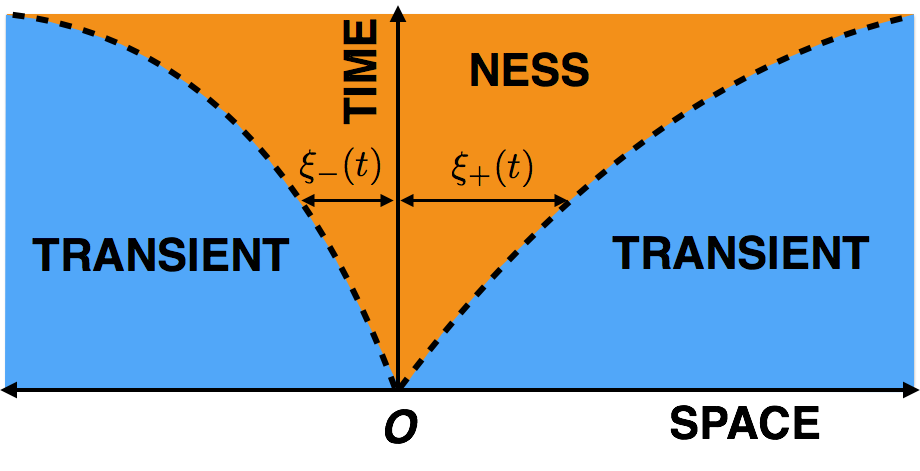}
\caption{\label{reset_dyn} (Color online) 
NESS gets established in a core region around the resetting center $O$
whose right and left frontiers $\xi_{\pm}(t)$ grow with time. Outside
the core region, the system is transient.}
\end{figure}

While the mechanism by which the stochastic resetting leads to an
eventual NESS has been well understood in the above studies, the
approach to the NESS in such systems is yet to be addressed. The goal
of this Rapid Communication is to study this temporal relaxation to
NESS in a wide variety of single particle, as well as many body
interacting systems such as fluctuating interfaces. It is useful first
to summarize our main results.  We compute exactly the time-dependent
probability distribution function (in presence of a resetting rate
$r$) $P_r(x,t)$ ---where $x$ may represent the position of a single
particle undergoing a stochastic motion with resetting to the origin,
or it may represent the height of a fluctuating interface at a fixed
point in space with resetting to the initial condition. As
$t\to \infty$, the system reaches the NESS characterized by
$P_r(x,\infty)$.  In systems without the $x\to -x$ symmetry, our
solution at finite $t$ shows that as time progresses, the NESS gets
established in an inner core region $[-\xi_{-}(t), \xi_+(t)]$ around
the point where the system is reset (which we take it to be the origin
$O$) ---outside this core regime, the system is still transient
(see \fref{reset_dyn}).  The frontiers $\xi_{\pm}(t)$, separating the
inner NESS regime from the outer transient regime, typically grow with
time as power laws, thus establishing NESS on larger and larger length
scales.  This phenomenon is conveniently captured by probing the PDF
on the scale $x\sim \xi_{\pm}(t)$, where it has the large deviation
form
\begin{equation}
P_r(x,t) \sim \exp\left[-t\, I \left(\frac{x}{\xi_{\pm}(t)}\right)\right]\,,
\label{ldf.0}
\end{equation}
where positive and negative fluctuations are scaled differently and
$I(y)$ represents the large deviation function (LDF).  For systems
with $x\to -x$ symmetry, there is only a single length scale
$\xi_+(t)=\xi_{-}(t)\equiv \xi(t)$ and $I(y)$ is symmetric.  The LDF
$I(y)$ characterizes the spatial dependence of the temporal
relaxation. At different points $x$, the system relaxes with a
different $x$ dependent rate. Alternatively, at a fixed point $x$, the
system undergoes a dynamical transition from a transient to the
stationary state at a characteristic time $t^*(x)$ such that
$\xi_{\pm} (t^*)=x$ (for positive and negative $x$ respectively).  We
compute $\xi_{\pm}(t)$ and $I(y)$ explicitly for several systems. We
show that $\xi_{\pm}(t)\sim t^{1/\nu_{\pm}}$ typically grow
algebraically at late times with a pair of new exponents. For the
underlying stochastic process without resetting the typical time
dependent length scale also grows algebraically $\sim t^{1/z}$ where
$z$ is the dynamical exponent, e.g., $z=2$ for simple diffusion. In
presence of resetting, we show that $\nu_\pm$ are generically smaller
than $z$. Furthermore, we show that the rate function $I(y)$ exhibits
a universal feature: it has a pair of singular points at $y=y_{+}^*$
and $y=-y_{-}^*$, corresponding to the frontiers of the core region
---signaling a dynamical phase transition. As discussed later,
physically, this LDF $I(y)$ and its singularity provides a sharp
spatio-temporal separation between typical and atypical trajectories
of the underlying stochastic process with resetting. It turns out
that, generically, the second derivative $I''(y)$ is discontinuous at
this pair of singular points.

{\em Diffusion of a single particle.}--- We begin with the
simple example of a single particle diffusing in one dimension
(generalization to higher dimensions is straightforward) whose
position is stochastically reset to a fixed position (which is taken
to be the origin without loss of generality) with a constant rate $r$.
Let $P_r(x,t)$ be the probability density function (PDF) of the
position $x$ of the particle at time $t$.  There arise two situations:
one in which no resetting events occur during the observation time $t$
(the probability of which being $e^{-r t}$) so that the particle moves
from the origin to $x$ as a free diffusion, and another in which the
last resetting event before $t$ occurs at time $t-\tau$ (and no
resetting occurs in the remaining time $\tau$ ---the PDF of which
being $re^{-r\tau}$) so that the particle moves as a free diffusion in
the final stretch of the time $\tau$. Therefore, taking into account
both situations (and integrating over $\tau$), one
gets~\cite{Evans:14}
\begin{equation}
P_r (x,t) = e^{-r t} P_0(x,t) + \int_0^t d\tau\, r e^{-r\tau}
P_0(x,\tau),
\label{master}
\end{equation}
where $P_0(x,t)=\exp[-x^2/(4Dt)]/\sqrt{4\pi Dt}$ is the propagator for
the particle to diffuse freely from the origin to the position $x$ in
time $t$ in the absence of the resetting. The NESS is obtained by
taking the $t\to \infty$ limit in \eref{master}, which yields~\cite{Evans:11}
\begin{equation}
P_r(x, t\to \infty)= \int_0^\infty d\tau\, r e^{-r\tau} P_0(x,\tau) =
\frac{\alpha}{2} \exp\bigl(-\alpha |x|\bigr),
\label{P_r-SS}
\end{equation}
where $\alpha=\sqrt{r/D}$. 
To analyse (\ref{master}) for finite $t$, it is convenient to reexpress
it using a change of variable $\tau=wt$, yielding
\begin{subequations}
\begin{equation}
P_r(x,t)= \frac{e^{-t \Phi(1,x/t)}}{\sqrt{4\pi D t}}  +
\frac{r\sqrt{t}}{\sqrt{4\pi
D}}\int_0^1 \frac{dw}{\sqrt{w}} \,e^{-t \Phi(w, x/t)}, 
\label{P_r-diffusion}
\end{equation} 
where
\begin{equation}
\Phi(w,y)=rw + \frac{y^2}{4 Dw}.
\label{action-diffusion}
\end{equation}
\end{subequations}

For large $t$ and fixed $y=x/t$, the integral in the second term can
be estimated by the saddle point method.  The function $\Phi(w,y)$
evidently has a single minimum with respect to $w$ at
$w^*=|y|/\sqrt{4Dr}$, obtained by setting
$\partial_w \Phi(w,y)|_{w=w^*}=0$. If $w^*<1$, the saddle point occurs
within the integration limits $w\in [0,1]$ and one gets,
from \eref{master} $P_r(x,t) \sim e^{-t\, \Phi(w^*,x/t)}$ for large
$t$, where $\Phi(w^*,y)= \alpha\, |y|$.  In contrast, for $w^*>1$, the
function $\Phi(w,y)$ has its lowest value in $w\in [0,1]$ at
$w=1$. Hence the integrand in the second term is dominated by the
regime at $w=1$ (and is of the same order as the first
term). Physically, this corresponds to trajectories which have
undergone zero (or almost zero) resettings up to time $t$.  One then
gets $P_r(x,t) \sim e^{-t\, \Phi(1,x/t)}$, with $\Phi(1,y)=
r+y^2/{(4D)}$. Summarizing, we obtain
\begin{subequations}
\label{LD-diffusion}
\begin{equation}
P_r(x,t)\sim e^{-t I \left(x/t\right)},
\label{LD0-diffusion}
\end{equation}
where the LDF 
\begin{equation}
I(y) = \begin{cases}\\[-2.2ex]\displaystyle
\alpha\, |y| & \text{for}~~|y| < y^*, \\[2mm]
\displaystyle
r + \frac{y^2}{4D} & \text{for}~~ |y| > y^*,\\[3mm]
\end{cases}
\label{I-diffusion}
\end{equation}
\end{subequations}
with $y^*=\sqrt{4Dr}$. 

\begin{figure}
\includegraphics[width=\hsize]{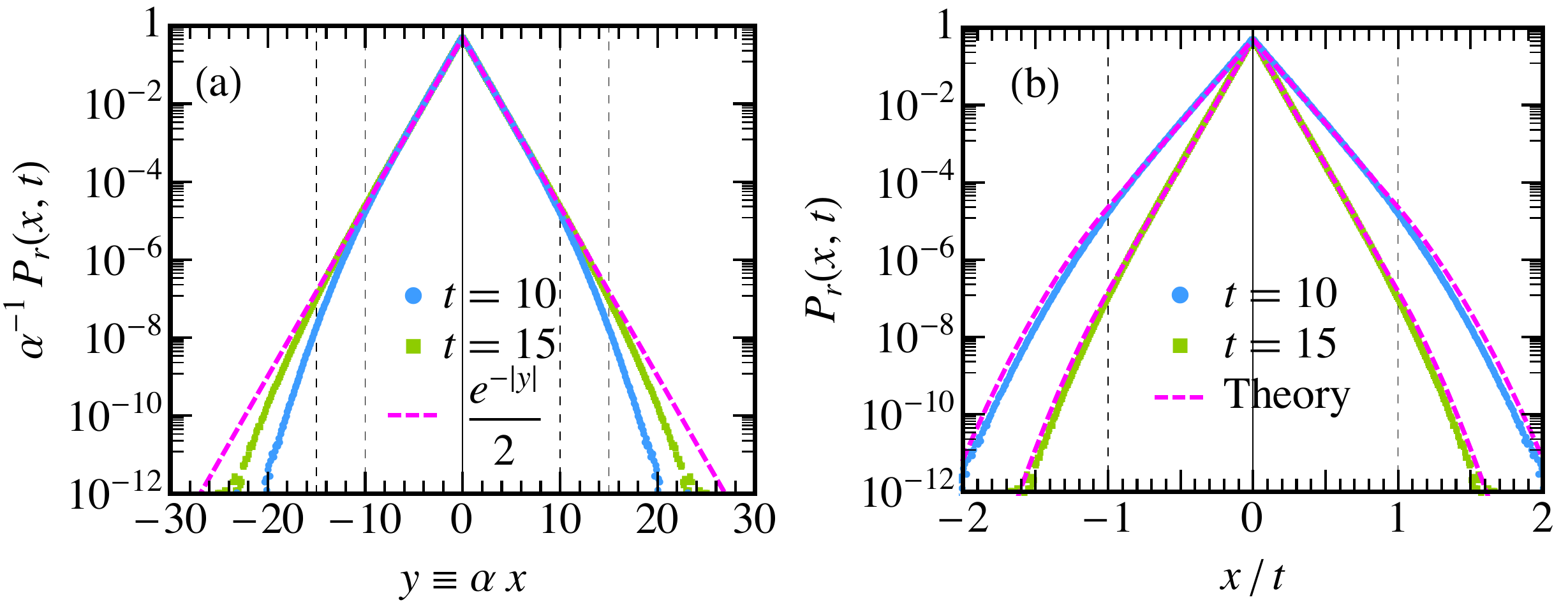}
\caption{\label{diffusion-fig} (Color online) PDFs of the position of
a particle diffusing in one dimension with a diffusion constant
$D=1/2$, whose position is stochastically reset to the origin at a
constant rate $r=1/2$. (a) The points are simulation data for the PDFs
at $t=10$ (blue circles) and $t=15$ (green squares).  The (magenta)
dashed line is the infinite time NESS given in \eref{P_r-SS}.  The
vertical dashed lines mark the positions $\pm y^*t$.  (b) Same data as
in (a) compared with large deviation result [\eref{LD-diffusion}] of
the PDF (normalized numerically), denoted by the (magenta) dashed
lines.  The dashed vertical lines at $y= \pm 1$ mark the $y^*$ at
which the LDF has a second order discontinuity. }
\end{figure}

Comparing \eref{LD0-diffusion} with \eref{ldf.0} shows that the
growing length scale $\xi(t) \sim t$, much larger than the typical
diffusion length scale $\sim \sqrt{t}$.  The linearity of the LDF for
$|y| < y^*$ implies that, for any large but finite $t$, there is an
interior spatial region $-y^*t < x < y^*t$, where NESS has been
achieved, as $P_r(x,t) \sim \exp(-\alpha |x|)$ becomes independent of
$t$ ---in agreement with \eref{P_r-SS}. However, there is still an
exterior region $|x| > y^*\, t$ that has not been relaxed to the NESS
yet. The boundaries between the two regions move at a constant speed
$y^*$.  From
\eref{I-diffusion}, it is easy to check that while $I(y)$ 
and its first derivative are both continuous at $y=\pm y^*$, its
second derivative has a discontinuity at $y=\pm y^*$.  Therefore, the
LDF has a second order discontinuity at the points $\pm
y^*$. \Fref{diffusion-fig} shows very good agreement between the above
large deviation form of the PDF and numerical simulations.  The above
analysis can be easily generalized to higher dimensional diffusion as
well as to other stochastic processes such as the fractional Brownian
motion (see Appendix).

What is the physical significance of this phase transition? The
probability density $P_r(x,t)$ can also be interpreted as the density
at time $t$ of a swarm of independent Brownian motions, each subjected
to stochastic resetting with rate $r$, all starting from the origin at
$t=0$. Our calculation shows that at time $t$ the density for
$|x|<y^* \, t$ becomes stationary, while is still time dependent for
$|x|>y^* \, t$. From the analysis above, it is clear that, for
$|x|>y^* \, t$, the density is typically of the form $\sim e^{-r\,t}
P_0(x,t)$ in Eq. (\ref{master}), i.e., it corresponds to particles
that have undergone almost no resetting up to time $t$.  This is of
course a very rare event and these particles in the outer region thus
have very atypical trajectories. In contrast, the particles in the
inner core region correspond to typical trajectories that have
undergone a large number of resettings -- leading to a stationary
behavior in this regime. The LDF $I(y)$ in Eq. (\ref{LD0-diffusion})
probes precisely the separation between these two regions, i.e.,
between the typical and the atypical trajectories. The singularity in
the LDF signifies a sharp separation between these two types of
particles. In any typical application of resetting, for instance in
the optimization of search algorithms, we would ideally like to keep,
at any given finite time $t$, only the typical trajectories and not
the atypical ones -- since the latter ones do not feel the resetting
at all. The LDF $I(y)$ and its associated singularity, that sharply
separates the two types of trajectories, thus provides a very useful
and practical way to select the typical ones at any given time
$t$. Even though we discuss it here in the context of a single
particle diffusion, it turns out that this picture is quite generic
and holds for arbitrary stochastic processes undergoing resetting and
even for spatially extended systems, such as fluctuating interfaces
that we discuss next.

{\em Fluctuating $(1+1)$-dimensional interfaces.}--- We
next look at the effect of resetting on extended correlated objects
such as a fluctuating interface growing with time over a linear
substrate of size $L$.  The interface is characterized by its height
field $H(x,t)$ which typically evolves via a stochastic
dynamics~\cite{Halpin:95,Barabasi:book,Krug:97}.  The height
fluctuation is measured by the relative height field, $h(x,t)=
H(x,t)-\overline{H(x,t)}$, where $\overline{H(x,t)}=L^{-1}\int_0^L
dx \, H(x,t)$ is the spatially averaged height.  The roughness of the
interface is usually measured by the interface width $W(L,t)$ defined
as $W^2(L,t) = \frac{1}{L} \int_0^L dx\, \langle h^2(x,t)\rangle$.
For a large class of interfaces, $W(L,t)$ increases with time as
$W(L,t)\sim t^\beta$ for $0\ll t\ll L^z$, before saturating to a
time-independent value $W_\mathrm{sat} \sim L^\alpha$ for $t \gg
L^z$. The scaling exponents $\alpha$, $\beta$ and $z$ are known as the
roughness exponent, growth exponent and dynamic exponent respectively,
and are related by the scaling relation $z=\alpha/\beta$, leaving only
two independent exponents~\cite{Halpin:95}. Moreover, in the growing
regime $0\ll t\ll L^z$, the full height distribution $P_0(h,t)$ has
the generic scaling form~\cite{Halpin:95,Barabasi:book,Krug:97}
\begin{equation}
P_0(h,t) \approx (\Gamma t)^{-\beta} g \bigl((\Gamma t)^{-\beta}h\bigr) \;,
\label{P_0-general}
\end{equation}
where the scaling function $g(x)$ is identical for all models
belonging to the same universality class while $\Gamma^{-1}$ is a
model-dependent microscopic time scale. For example, for simple linear
stochastic interface models belonging to the Edwards-Wilkinson (EW)
universality class (where $\beta=1/4$ and $z=2$), the height
distribution at all times is simply Gaussian~\cite{EW:82} with
$g(x)\propto \exp[-x^2]$. Another widely studied class of growing
interfaces in $(1+1)$ dimensions belong to the Kardar-Parisi-Zhang
(KPZ) universality
class~\cite{KPZ:86,Halpin:95,Barabasi:book,Krug:97}. In this case one
has $\beta = 1/3$ and $z=3/2$ while the scaling function $g(x)$, for a
flat initial condition $H(x,t=0) = 0$, is related to the Tracy-Widom
distribution \cite{TW} associated to the Gaussian Orthogonal Ensemble
(GOE) of random matrices \cite{KPZ:solution}, $f_1(\chi)$, which
describes the fluctuations of the largest eigenvalue of GOE
matrices. One has indeed $g(x) = f_1(x+\langle \chi \rangle)$ where
$\langle \chi \rangle = \int_{-\infty}^\infty d\chi f_1(\chi)$. While
the full form of $f_1(\chi)$ is rather nontrivial--- the tails have
simpler non-Gaussian forms: $f_1(\chi) \sim \exp(-|\chi|^3/24)$ as
$\chi\to -\infty$ and $f_1(\chi)\sim \exp(-2\chi^{3/2}/3)$ as
$\chi\to\infty$.

Let us now consider the height field of such a generic $(1+1)$
dimensional interface evolving under its own dynamics and subject it
to resetting to its initial height profile at constant rate $r$.
Following our discussion prior to \eref{master}, one can relate the
height distribution $P_r(h,t)$ in the presence of resetting to that of
$P_0(h,t)$ without resetting via the same equation \eref{master}, with
$x$ replaced by $h$. Using the scaling form for $P_0(h,t)$ in
\eref{P_0-general} and making the change of variable
$\tau=wt$, \eref{master} reduces to
\begin{align}
P_r(h,t)&\approx (\Gamma t)^{-\beta} e^{-r t} g \bigl((\Gamma
t)^{-\beta}h\bigr)\notag\\
&+ rt (\Gamma t)^{-\beta} \int_0^1 dw\, w^{-\beta} e^{-rt w}
g \bigl((\Gamma
t)^{-\beta} h w^{-\beta}\bigr).
\label{master-interface}
\end{align}
As before, $P_r(h,t)$ approaches a stationary distribution as
$t\to \infty$ for any $r>0$~\cite{Gupta:14}
\begin{subequations}
\label{interface-NESS}
\begin{equation}
P_r(h, t\to \infty)\approx(\Gamma/r)^{-\beta}
G_\beta\bigl((\Gamma/r)^{-\beta} h\bigr), 
\end{equation}
where the scaling function is given by
\begin{equation}
G_\beta(x)= \int_0^\infty dy\, y^{-\beta} e^{-y}
g\bigl(x y^{-\beta} \bigr).
\end{equation}
\end{subequations}

Now to investigate the approach to the NESS, we consider the generic
case when $g(x)\sim \exp(-a_\pm |x|^{\gamma_\pm})$ as
$x\to \pm \infty$. For example, for the KPZ with flat initial condition,
$\gamma_+=3/2$, $a_+=2/3$ and $\gamma_{-}=3$, $a_{-}=1/{24}$.
Substituting these generic tails of $g(x)$ in 
\eref{master-interface} we obtain that the large
deviation form of $P_r(h,t)$ (see Appendix)
\begin{subequations}
\begin{equation}
P_r(h,t) \sim e^{-t\, I(h\, t^{-1/\nu_\pm})},
\label{prob_height} 
\end{equation}
where the LDF is given by
\begin{equation}
I(y) = \begin{cases}\displaystyle
\frac{r\,|y|^{\nu_\pm}}{\beta\nu_\pm (y_\pm^*)^{\nu_\pm}}\,
 & \text{for}~~|y| < y_\pm^*, \\[3mm] 
\displaystyle
r + b_\pm |y|^{\gamma_\pm} & \text{for}~~ |y| > y_\pm^* \;.
\end{cases}
\label{I-interface}
\end{equation}
\end{subequations}
The exponents $\nu_{\pm} = \gamma_{\pm}/(1+\beta \gamma_{\pm})$ and
the boundaries $y^*_{\pm}$ as well as the constants $b_{\pm}$ can be
explicitly computed (see Appendix). The signs $\pm$
are chosen for $y>0$ and $y<0$
respectively. Comparing \eref{prob_height} and \eref{ldf.0}, we see
that the growing scale, separating the inner NESS regime from the
outer transient regime in the height space, $\xi_{\pm}(t) \sim
t^{1/\nu_\pm}$ is asymmetric (different respectively for positive and
negative height fluctuations).  The height fluctuations in the
intermediate range $-y_-^* t^{1/\nu_-} < h < y_+^* t^{1/\nu_+}$
reaches NESS. Moreover, $I(y)$ again has a second order discontinuity
at the two singular points $y_\pm^*$.

\begin{figure}
\includegraphics[width=\hsize]{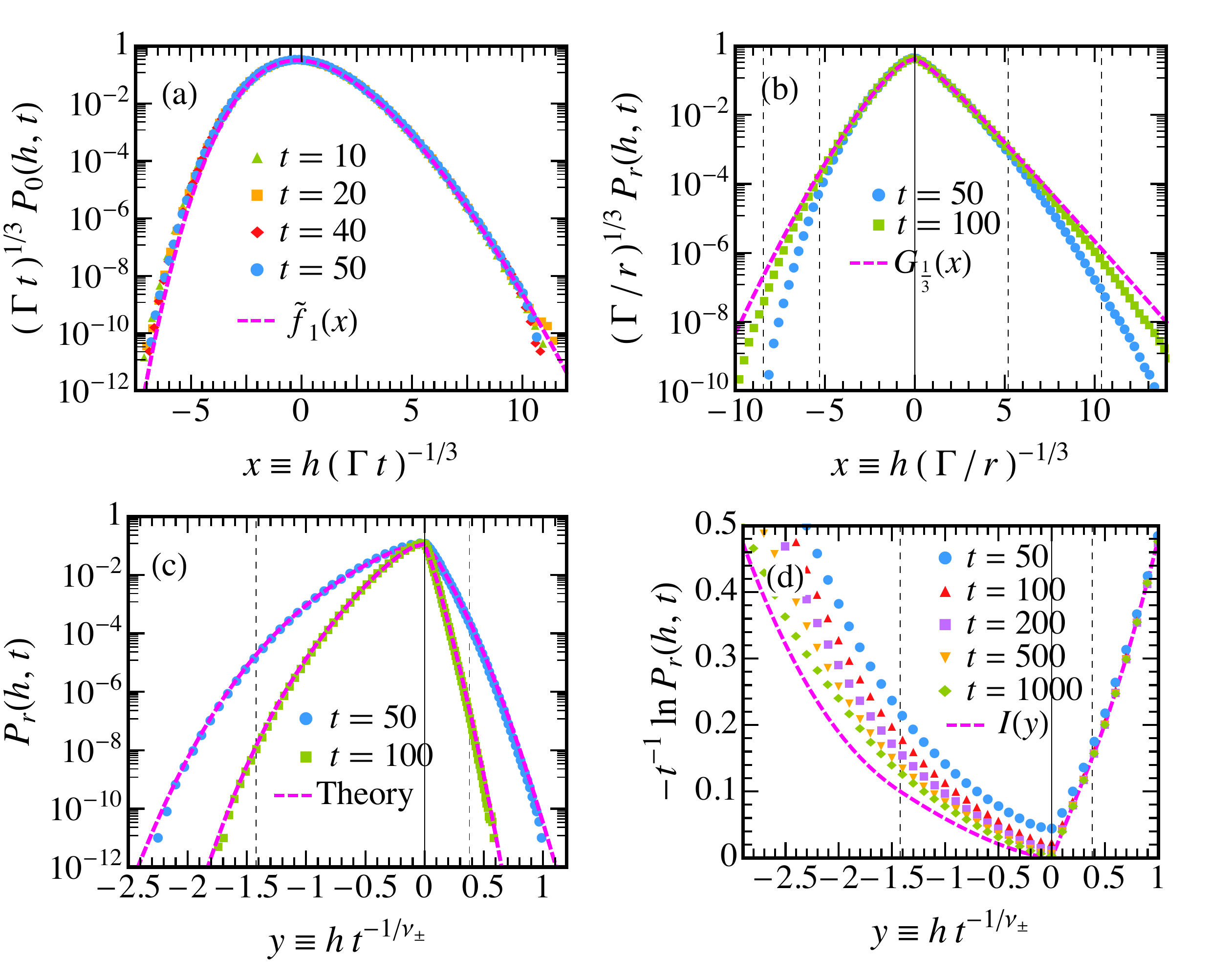}
\caption{\label{interface-plot}(Color online). (a) The PDFs of the relative
heights of a periodic interface of size $L=2^{15}$ at various times,
generated from a TASEP 
(shown by points) are collapsed by choosing $\Gamma=2.4$ to the TW GOE
PDF with zero mean, shown by the (magenta) dashed line. (b) The points
are simulation data for the PDF of the relative heights at $t=50$
(blue circles) and $t=100$ (green squares), with resetting with rate
$r=0.05$.  The (magenta) dashed line is the infinite time NESS given
in \eref{interface-NESS}. The vertical dashed lines mark the positions
$x^*_\pm =\pm y^*_\pm t^{1/\nu_\pm}$. (c) Same data as in (b) compared
with \eref{master-interface} denoted by the (magenta) dashed
lines. The vertical dashed lines mark the positions $\pm y^*_\pm$.
(d) The LDFs computed numerically from (\ref{master-interface}) for
various $t$ are compared with (\ref{I-interface}).}
\end{figure}

To perform simulations, we consider an interface model generated from
the well known totally asymmetric simple exclusion process
(TASEP)~\cite{Liggett:book} on a ring. This interface model
is known to belong to the KPZ class~\cite{Halpin:95,Krug:97}.
\Fref{interface-plot}~(a) shows that the scaled PDF for
various times in the growing regime ($1 \ll t \ll L^z$) can be
collapsed by choosing an unique value of $\Gamma$ (same for all $t$),
to the TW GOE PDF with the mean shifted to zero. We use this value of
$\Gamma$ subsequently. We next simulate the dynamics of the interface
in presence of resetting (with rate $r$) to the flat initial
condition.  It is evident from \fref{interface-plot}~(b) that at long
times the central part $|h| \ll t^{1/\nu_\pm}$ of the distribution of
the relative heights reaches the NESS given
by \eref{interface-NESS}. \Fref{interface-plot}~(c) shows very good
agreement between the simulation and the theoretical results obtained
from numerical integration
of \eref{master-interface}. In \fref{interface-plot}~(d) we plot the
LDFs computed numerically from (\ref{master-interface}) for various
$t$, which converges to (\ref{I-interface}) as time increases.

{\em Discussion.}--- In summary, we have studied analytically the
approach to the stationary state of several systems subjected to
resetting to the initial condition at a constant rate.  We have shown
that the relaxation mechanism in these systems is highly unusual.  In
typical systems approaching to their stationary states, the late time
relaxation is governed by a single time scale independent of space. In
systems with resetting studied here, the late time relaxation is space
dependent and is characterized by the LDF $I(x/\xi(t))$ defined
in \eref{ldf.0}. The growing length (or height) scale around the
resetting center behaves at late times as $\xi(t)\sim t^{1/\nu}$ where
$\nu < z$ is a new exponent typically smaller than the dynamical
exponent $z$ of the process without resetting. For system without the
$x \to -x$ symmetry, such as in the KPZ equation, there are actually a
pair of asymmetric length scales $\xi_{\pm}(t) \sim t^{1/\nu_{\pm}}$
where $\nu_{\pm} < z$. We have computed this LDF $I(y)$ explicitly in
several examples and found that it exhibits a universal feature: it
has a a pair of singular points where the second derivative is
discontinuous. The singularities in $I(y)$ provide a sharp separation
between typical and atypical trajectories and thus can be useful in
various applications such as search optimization using resetting, in
detecting and getting rid of the undesirable atypical trajectories
(that have not undergone resetting) at any finite time $t$. The
important point is the presence of these singularities providing a
sharp separation -- the actual order of the singularities is of less
relevance.  We conclude by noting that singularities in the LDFs have
also been found recently in several different contexts and there is a
growing interest in understanding the significance of these
singularities~\cite{Touchette:09,Sabhapandit:11,Kumar:11,Sabhapandit:12,Bunin:12,Pal:13,MS:14}. In
this paper we have provided a clear physical meaning of these
singularities that act like separatrices between typical and atypical
trajectories.

{\em Acknowledgment.}---
We acknowledge support of the Indo-French Centre for the Promotion of
Advanced Research under Project No. 4604-3.

{\em Appendix.}--- In this appendix, we give some details of the
calculations described in the main text.

{\em Fractional Brownian Motion.}--- The analysis of $P_r(x,t)$
carried out in the main text above for the simple Brownian motion can
be easily generalized to the fraction Brownian motion (fBM), which
represents a Gaussian process with zero mean and two-time correlator,
$\langle x(t_1)x(t_2\rangle= t_1^{2H} + t_2^{2H}-|t_1-t_2|^{2H}$ where
$0<H<1$ represent the Hurst index. Clearly $H=1/2$ corresponds to the
simple Brownian motion.  For simplicity we discuss the one dimensional
case, though generalization to higher dimensions is straightforward.
The propagator for the fBM in one dimension, starting initially at the
origin, again has the simple Gaussian form
\begin{equation}
P_0(x,t) = \frac{1}{ \sqrt{4\pi Dt^{2H}}}\, \exp\left[-\frac{x^2}{4\,D\,t^{2H}}\right]\,.
\label{fbm.1}
\end{equation}
We then subject the particle undergoing fBM to resetting at the origin
with a constant rate $r$.  Then, the propagator $P_r(x,t)$ in presence
of the resetting satisfies the generic relation given
by \eref{master}.
Substituting the bare propagator $P_0(x,t)$ from \eref{fbm.1} in
\eref{master}, we can then carry out the same analysis as the
$H=1/2$ case in the main manuscript.  As in the case of $H=1/2$, it
turns out that for general $0<H<1$, the growing length scale
$\xi(t)\sim t^{H+1/2}$, much larger than the typical spread $\sim
t^{H}$ for large $t$. In the scaling limit $x\to \infty$, $t\to
\infty$ with $x/t^{H+1/2}$ fixed, we obtain, using the saddle point
method (discussed already for $H=1/2$ in the main manuscript)
\begin{subequations}
\begin{equation}
P_r(x,t)\sim e^{-t\, I \left(x/t^{H+1/2}\right)},
\label{P_r-FBM}
\end{equation}
where the large deviation function (LDF) is
\begin{align}
&\qquad\qquad
I(y) = \begin{cases}\\[-2.2ex]\displaystyle
\alpha_H\,
|y|^\frac{1}{H+1/2} & \text{for}~~|y| < y^*, \\[2mm]
\displaystyle
r + \frac{y^2}{4D} & \text{for}~~ |y| > y^*,\\[3mm]
\end{cases}\\
\label{I-FBM}
&\text{with}~\alpha_H=
r\left(1+\frac{1}{2H}\right)\left(\frac{H}{2Dr}\right)^{\frac{1}{2H+1}}
~\text{and}~ y^*=\sqrt{\frac{2Dr}{H}}. \notag
\end{align}
\end{subequations}
It is easy to check that both the LDF and its first derivative are
continuous across $\pm y^*$, while the second derivative has a discontinuity
at $\pm y^*$. The NESS is realized in an interior region where
$P_r(x,t)\sim \exp\bigl(-\alpha_H\, |x|^\frac{1}{H+1/2}\bigr)$ becomes
independent of time, with the boundaries between this NESS and the
outer transient regime moving out with time as $|x^*|=y^* t^{H+1/2}$.


{\em The case of the Kardar-Parisi-Zhang equation.}--- A widely
studied class of growing interfaces in $(1+1)$ dimensions belong to
the Kardar-Parisi-Zhang (KPZ) universality
class~\cite{Halpin:95,Barabasi:book,Krug:97}.  Here the height field
evolves via the nonlinear KPZ equation~\cite{KPZ:86}
\begin{equation}\label{eq:KPZ}
\frac{\partial H}{\partial t}=\nu \frac{\partial^2H}{\partial x^2} 
+ \frac{\lambda}{2} \left(\frac{\partial H}{\partial x} \right)^2
+\eta(x,t), 
\end{equation}
where $\nu$ is the surface tension, $\lambda$ represents the strength
of the nonlinearity, and $\eta(x,t)$ is a Gaussian white noise with
zero mean and correlations $\langle \eta(x,t) \eta(x',t')\rangle =
2D \delta(x-x') \delta (t-t')$.  In absence of nonlinearity ($\lambda=0$),
the KPZ equation (\ref{eq:KPZ}) reduces to the EW equation.

The scaling exponents for the KPZ equation in one dimension are well
known~\cite{Halpin:95}: $\alpha=1/2$, $\beta=1/3$ and $z=3/2$. In the
KPZ case, the spatially averaged height $\overline{H(x,t)} = L^{-1}
\int_0^L dx\,H(x,t)$ grows linearly with time with a non-zero velocity
$v_\infty=(\lambda/2)L^{-1}\int_0^L \langle(\partial H/\partial x)^2
\rangle\, dx$. In contrast, for the EW case, $\overline{H(x,t)}\sim
\sqrt{t/L}$ for large $t$.  In the growing regime $0\ll t\ll L^z$ to
which we focus below, the PDF $P_0(h,t)$, while trivially Gaussian for
the EW case, is highly nontrivial in the KPZ case and has been a
subject of intense investigations in recent times.  It has been solved
exactly for the KPZ equation only very
recently~\cite{KPZ:solution}. It turns out to depend on the initial
condition of the height profile~\cite{KPZ:solution,halpin:2014}.

For example, for a {\em flat initial profile}, the height $H(x,t)$ of
the interface can be written as
\begin{equation}
H(x,t)=v_\infty\, t +  (\Gamma t)^{1/3} \chi(x),
\end{equation}
where $\Gamma$ is a constant that depends on the parameters of the
interface model, and $\chi$ is a time independent random variable
distributed according to the so-called Tracy-Widom (TW) distribution
$f_1(\chi)$, that characterizes the fluctuations of the largest eigenvalue
of random matrices in the Gaussian orthogonal ensemble
(GOE) \cite{TW}. While the full form of $f_1(\chi)$ is rather nontrivial---
the tails have simpler non-Gaussian forms:
$f_1(\chi) \sim \exp(-|\chi|^3/24)$ as $\chi\to -\infty$ and
$f_1(\chi)\sim \exp(-2\chi^{3/2}/3)$ as $\chi\to\infty$. In terms of
$\chi$, we have
\begin{equation}
h(x,t)=(\Gamma t)^{1/3} \bigl[\chi(x)- \overline{\chi}\bigr],
\end{equation}
where $\overline{\chi}=L^{-1}\int_0^L \chi(x) \, dx$. The law of large
number dictates that $\overline{\chi}\to \langle \chi \rangle$ in the
limit $L\to \infty$, so that $\langle h \rangle =0$. Therefore, in
this case, in the scaling limit of $t\to\infty$, $h\to\infty$ while
keeping $h/t^{1/3}$ fixed, the height fluctuation is distributed
according to
\begin{equation}
P_0(h,t) \approx (\Gamma t)^{-1/3} \widetilde{f}_1\bigl( (\Gamma
t)^{-1/3}h\bigr) \;,
\end{equation}
where $\widetilde{f}_1(x)=f_1\bigl(x+\langle \chi \rangle\bigr)$. This
is indeed of the form announced in \eref{P_0-general} in the main text
above with $\beta = 1/3$ and $g(x) = \widetilde{f}_1(x)$.  Similarly,
for the droplet initial configuration (curved geometry), the scaling
function for the height distribution is the TW distribution
corresponding to the Gaussian unitary ensemble (GUE).

{\em Generic interfaces: saddle point calculation.}--- Let us now
analyze the height field of a generic $(1+1)$ dimensional interface
evolving under its own dynamics and subject it to resetting to its
initial height profile at constant rate $r$.  In this case, the full
height distribution $P_0(h,t)$, in the absence of resetting ($r=0$),
has the generic scaling form~\cite{Halpin:95,Barabasi:book,Krug:97}
given by \eref{P_0-general}.  As shown in the main text, one can
relate the height distribution $P_r(h,t)$ in the presence of resetting
to that of $P_0(h,t)$ without resetting via the \eref{master} with $x$
replaced by $h$. Using the scaling form for $P_0(h,t)$ in
(\ref{P_0-general}) and making the change of variable $\tau=wt$, it
reduces to
\begin{align}
P_r(h,t)&\approx (\Gamma t)^{-\beta} e^{-r t} g \bigl((\Gamma
t)^{-\beta}h\bigr) \notag\\ +& rt (\Gamma t)^{-\beta} \int_0^1 dw\,
w^{-\beta} e^{-rt w} g \bigl((\Gamma t)^{-\beta} h w^{-\beta}\bigr).
\label{master-interface}
\end{align}
In this case, as in the case of single particle diffusion, $P_r(h,t)$
approaches a stationary distribution as $t\to \infty$ for any $r>0$,
given by~\cite{Gupta:14}
\begin{subequations}
\label{interface-NESS}
\begin{equation}
P_r(h, t\to \infty)\approx(\Gamma/r)^{-\beta}
G_\beta\bigl((\Gamma/r)^{-\beta} h\bigr), 
\end{equation}
where the scaling function is given by
\begin{equation}
G_\beta(x)= \int_0^\infty dy\, y^{-\beta} e^{-y}
g\bigl(x y^{-\beta} \bigr).
\end{equation}
\end{subequations}

Now to investigate the approach to the NESS, we consider the generic
case when $g(x)\sim \exp(-a_\pm |x|^{\gamma_\pm})$ as
$x\to \pm \infty$. For example, for the KPZ with flat initial condition,
$\gamma_+=3/2$, $a_+=2/3$ and $\gamma_{-}=3$, $a_{-}=1/{24}$.
Substituting these generic tails of $g(x)$ in 
\eref{master-interface} we obtain
\begin{subequations}
\begin{align}
P_r(h,t)&\sim (\Gamma t)^{-\beta} e^{-t \Phi(1,ht^{-1/\nu_\pm})}
\notag\\
&+ rt (\Gamma t)^{-\beta} \int_0^1 dw\,w^{-\beta} e^{-t \Phi(w,ht^{-1/\nu_\pm})} \;,
\label{P_r-interface}
\end{align}
where $\nu_\pm = \gamma_\pm/(1+\beta\gamma_\pm)$ and
\begin{equation}
\Phi(w,y)= rw + \frac{b_\pm|y|^{\gamma_\pm}}{ w^{\beta\gamma_\pm}}
\quad \text{with}
~~b_\pm=\frac{a_\pm}{\Gamma^{\beta\gamma_\pm}}.
\label{action-interface}
\end{equation}
\end{subequations}
With respect to the variable $w$, the function $\Phi(w,y)$ has a
unique minimum at $w^*=\bigl(|y|/y_\pm^*\bigr)^{\nu_\pm}$, where
\begin{math}\displaystyle
y_\pm^*= \bigl(r/[\beta\gamma_\pm b_\pm]\bigr)^{1/\gamma_\pm}.
\end{math}
As in the case of single particle diffusion discussed after Eq. (\ref{action-diffusion}), if $w^*<1$, the
most dominant contribution to \eref{P_r-interface} comes from the
neighborhood of $w^*$ so that $P_r(h,t)\sim e^{-t
  \Phi(w^*,h\,t^{-1/\nu_\pm})}$, whereas for $w^* >1$, it is dominated
by the boundary terms so that $P_r(h,t)\sim e^{-t
  \Phi(1,h\,t^{-1/\nu_\pm})}$. Therefore, the large deviation form of
$P_r(h,t)$ is given by
\begin{subequations}
\begin{equation}
P_r(h,t) \sim e^{-t\, I(h\, t^{-1/\nu_\pm})},
\label{prob_height} 
\end{equation}
where the LDF is given by 
\begin{equation}
I(y) = \begin{cases}\displaystyle
\frac{r\,|y|^{\nu_\pm}}{\beta\nu_\pm (y_\pm^*)^{\nu_\pm}}\,
 & \text{for}~~|y| < y_\pm^*, \\[3mm] 
\displaystyle
r + b_\pm |y|^{\gamma_\pm} & \text{for}~~ |y| > y_\pm^*,
\end{cases}
\label{I-interface}
\end{equation}
\end{subequations}
where the $\pm$ signs are chosen for $y>0$ and $y<0$
respectively. This yields the expression given in Eq. (9b), together
with the explicit expression for the boundaries $y^*_{\pm}$ and for
the constants $b_{\pm}$ [see Eq. (\ref{action-interface}) and below].
Moreover, one can show that $I(y)$ in (\ref{I-interface}) has a second
order discontinuity at the two singular points $y_\pm^*$.

{\em Generic second order discontinuity in the large deviation
function.}--- We now consider a generic system described by a
stochastic variable $x$ (which may represent the position of a
particle undergoing generic stochastic dynamics, or may represent the
height of a fluctuating interface). Let $P_0(x,t)$ be the bare
probability density function (PDF) of $x$ at time $t$, starting at
$x=0$, in the absence of resetting.  Let $P_r(x,t)$ denote the PDF of
$x$ at time $t$ in the presence of resetting to the initial condition
$x=0$ at a constant rate $r$. The PDF $P_r(x,t)$ is related to the
bare propagator $P_0(x,t)$ via the general relation in \eref{master}.
In the limit $t\to \infty$, $P_r(x,t)$ in \eref{master} approches the
stationary distribution
\begin{equation}
P_r(x, t\to \infty)= r\, \int_0^{\infty} d\tau\, e^{-r\, \tau}\, P_0(x,\tau)\, ,
\label{stat.1}
\end{equation}
which is fully determined by the bare PDF $P_0(x,t)$. 

In order to study the approach to the stationary state at late times,
we need to analyze \eref{master} for finite but large $t$. To proceed,
it is convenient to make a change of variable $\tau=w\,t$ in the
integral which yields
\begin{equation}
P_r(x,t)= e^{-r\,t}\, P_0(x,t)+ r\,t\, \int_0^1 dw\, e^{-t\, r\, w}\,
P_0(x, w\, t)\,.
\label{master1_supp}
\end{equation}
We are interested in the behavior of $P_r(x,t)$ for large $x$ and
large $t$. Hence, inside the integral we need to substitute the large
$x$, large $t$ behavior (with $w$ fixed) of $P_0(x, w\, t)$.  For
generic self-affine systems where $x(t) \sim t^\beta$, the PDF
$P_0(x,t)$ is expected to have a scaling form, for large $x$, large
$t$, keeping $x/t^{\beta}$ fixed
\begin{equation}
P_0(x,t) \approx \frac{1}{(\Gamma\, t)^{\beta}}\, g \left(\frac{x}{(\Gamma\, t)^{\beta}}\right) \;,
\label{P_0-general_supp}
\end{equation} 
where $g(y)$ is the scaling function and $\Gamma^{-1}$ is a
model-dependent microscopic time scale, not important for our
analysis, and henceforth set to unity without any loss of
generality. Substituting the scaling form \eref{P_0-general_supp} of
$P_0(x,t)$ in \eref{master1_supp} gives
\begin{align}
P_r(x,t)&= t^{-\beta}\,e^{-r\,t}\,g\left(\frac{x}{t^{\beta}}\right) 
\notag\\&+ r\,t^{1-\beta}\, \int_0^1 \frac{dw} 
{w^{\beta}}\,e^{-t\, r\, w}\, g\left(\frac{x}{(w\,t)^{\beta}}\right).
\label{master2_supp}
\end{align}

Next, we anticipate that $P_r(x,t)$ will have a large deviation form 
[see \eref{ldf.0}]
\begin{equation}
P_r(x,t) \sim \exp\left(- t\, I\left(\frac{x}{\xi(t)}\right)\right)\,,
\label{ldf_supp}
\end{equation}
where $I(y)$ is the rate function and $\xi(t)$ represents the growing
length scale associated with atypically large fluctuations of $x$ that
are much bigger than the typical fluctuations $x\sim t^{\beta}$.  Note
that in some cases (such as for the fluctuating interfaces belonging
to the KPZ universality class), the $x\to -x$ symmetry is broken. In
such cases, the positive and negative large fluctuations of $x$ occur
at different length scales $\xi_{\pm}(t)$ and accordingly,
$P_r(x,t)\sim \exp[-t\, I(x/\xi_{\pm}(t))]$ for positive and negative
fluctuations respectively [see \eref{ldf.0}].  In these systems, we
then need to scale the positive and negative fluctuations differently.
However, in problem with $x\to -x$ symmetry, there is only a single
scale $\xi(t)$ and the LDF $I(y)$ is symmetric. For simplicity, we
restrict ourselves below only to the positive fluctuations, as the
analysis of the negative fluctuations is similar.

In order to probe the behavior of $P_r(x,t)$ on the scale $x\sim
\xi(t)\gg t^{\beta}$, we then need to substitute the tail behavior of
$g(y)$ for large $y$ in the integrand of the second term in
\eref{master2_supp}. Generically, $g(y) \sim \exp[- a\, y^{\gamma}]$
for large positive $y$, where $a$ is an unimportant model dependent
constant. Substituting this tail behavior in the integrand in
\eref{master2_supp} we get
\begin{align}
P_r(x,t) &\sim  t^{-\beta} e^{-t \Phi\left(1,x/\xi(t)\right)} 
\notag\\&+ r\, t^{1-\beta}\,\int_0^1 dw\,w^{-\beta} e^{-t 
\Phi\left(w,x/\xi(t)\right)},
\label{master3_supp}
\end{align}
where $\xi(t)= t^{\beta+1/\gamma} \gg t^{\beta}$ and 
\begin{equation}
\Phi(w,y) = r\, w+ \frac{a\,y^{\gamma}}{w^{ \gamma\, \beta}}\, .
\label{phi_supp}
\end{equation}
Evidently $\Phi(w,y)$, as a function of $w$ (but fixed $y$), has a unique minimum at
$w^*(y)$, determined from 
\begin{equation}
\partial_w\Phi(w,y)\Big|_{w=w^*}=0 \quad\quad {\rm where}\quad \partial_w\equiv \frac{\partial}{\partial w}\, .
\label{min_supp}
\end{equation}
Actually the subsequent analysis will be very general and we do not
need to use the specific form of $\Phi(w,y)$ ---the only fact we will
use is that $\Phi(w,y)$ has a unique minimum at $w^*(y)$, determined
via the minimization in \eref{min_supp}.

\begin{figure}
\includegraphics[width=\hsize]{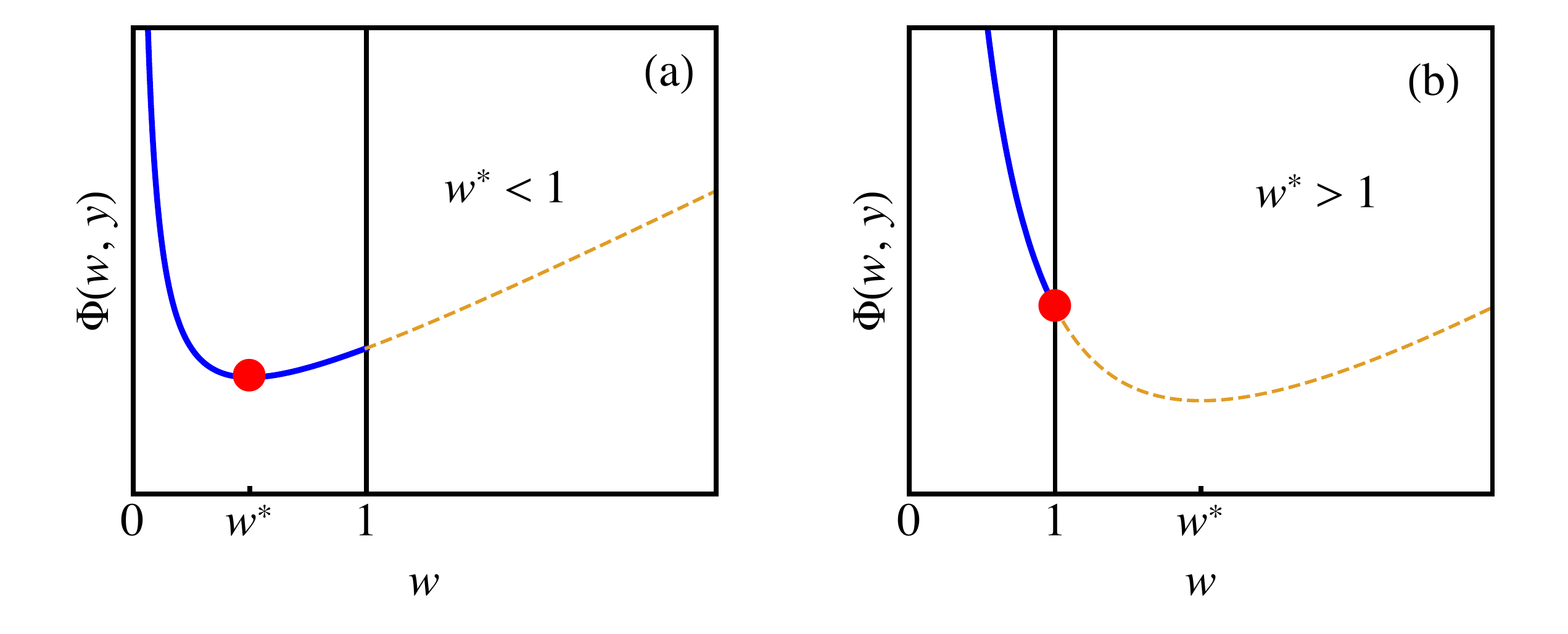}
\caption{\label{Phi-plot} Schematic plot of the function $\Phi(w,y)$
  as a function of $w$ (but fixed $y$), which has a unique minimum at
  $w^*(y)$, determined from \eref{min_supp} for the two cases (a) $w^*
  <1$ and (b) $w^* >1$. For the case (a), the integral in
  \eref{master3_supp} is dominated by the minimum $w^*$, whereas for
  the case (b) the integral is dominated by the upper limit $w=1$ of
  the integral. The minimum values of $\Phi(w,y)$ in the interval
  $w\in [0,1]$ in both cases, are marked by the red circles.}
\end{figure}

For large $t$, the integral in \eref{master3_supp} is dominated by the
minimum $w^*(y)$, as long as $w^*(y)<1$, i.e., the saddle point lies
within the limits of the integral $w\in [0,1]$, as shown in
\fref{Phi-plot}. For $w^*(y)>1$, the integral is dominated by the
upper limit $w=1$ of the integral (see \fref{Phi-plot}), which is of
the same order as the first term.  Thus, we have a critical point at
$y=y^*$ which is determined from $w^*(y^*)=1$. Thus evaluating the
integral in \eref{master3_supp} by the saddle point method for large
$t$ gives the large deviation form given in \eref{ldf_supp} where
\begin{equation}
I(y) = \begin{cases}\\[-2.2ex]\displaystyle
\Phi(w^*(y),y) & \text{for}~~y < y^*, \\[2mm]
\displaystyle
\Phi(1,y) & \text{for}~~ y >y^* \;. \\[3mm]
\end{cases}
\label{I_supp}
\end{equation}
This is the generic mechanism of the dynamical transition---the rate
function $I(y)$ changes its behavior at $y= y^*$. While the function
$I(y)$ is evidently continuous at $y=y^*$, we show below that while
its first derivative $I'(y)$ is also continuous at $y=y^*$, the second
derivative $I''(y)$ is discontinuous, signaling a second order
dynamical transition.

Consider first a function of $y$ of the form $H(y)= \Phi(u(y),y)$
where the first argument $u(y)$ depends implicitly on $y$. The
derivatives of $H(y)$ can be easily determined by the chain rule.  For
example, the first derivative is given by
\begin{equation}
H'(y) = \frac{dH}{dy}= \partial_u \Phi(u,y)\, u'(y) + \partial_y \Phi(u,y)\,.
\label{h1_supp}
\end{equation}
Similarly, the second derivative is given by
\begin{align}
H''(y) &= \partial_u^2 \Phi(u,y)\, \left[u'(y)\right]^2 +2\, \partial_u\partial_y \Phi(u,y)\, u'(y) \notag\\
&+ \partial_u \Phi(u,y)\, u''(y) + \partial_y^2 \Phi(u,y)\,.
\label{h2_supp}
\end{align}

Now, let us first consider the case $y<y^*$ in \eref{I_supp}. To
evaluate $I'(y)$, we substitute $u(y)=w^*(y)$ in \eref{h1_supp}. Using
the minimization condition in \eref{min_supp}, we obtain
\begin{equation}
I'(y) = \partial_y \Phi(w^*(y),y)\quad\quad {\rm for}\quad y<y^* \,.
\label{I1_supp}
\end{equation}
If now $y\to y^*$ from below, using $w^*(y^*)=1$, we get
\begin{equation}
I'(y\to y^*)= \partial_y \Phi(1,y)\Big|_{y=y^*}\, .
\label{h11_supp}
\end{equation} 
For $ y> y^*$ from above, we have $I(y)= \Phi(1,y)$ from \eref{I_supp}. Hence as $y\to y^*$ from above,
$I'(y\to y^*) = \partial_y \Phi(1,y)\Big|_{y=y^*}$. Comparing this with \eqref{h11_supp}, we see that
$I'(y)$ is continuous at $y=y^*$.

We next consider the second derivative $I''(y)$ as $y\to y^*$ from
below and above.  Consider first the case $y<y^*$.  Substituting
$u(y)= w^*(y)$ in \eqref{h2_supp} and using the minimization condition
\eref{min_supp}, we get
\begin{align}
I''(y) &= \partial_{w^*}^2 \Phi(w^*,y)\, \left[(w^*)'(y)\right]^2 \notag\\&+2\, \partial_{w^*}\partial_y \Phi(w^*,y)\, (w^*)'(y)
+ \partial_y^2 \Phi(w^*,y)\,.
\label{h21_supp}
\end{align}
Now, as $y\to y^*$ from below, $w^*(y) \to 1$ and we get
\begin{align}
I''(y\to y^*) =& \Bigl[\partial_{w^*}^2 \Phi(w^*,y)\,
  \left[(w^*)'(y)\right]^2 \notag\\
&~+2\, \partial_{w^*}\partial_y 
\Phi(w^*,y)\, 
(w^*)'(y)\Bigr]\Big|_{w^*=1, y=y^*} \notag\\
&+ \partial_y^2 \Phi(1,y)\Big|_{y=y^*}\, . 
\label{h22_supp}
\end{align}
In contrast, for $y>y^*$, $I(y)= \Phi(1,y)$ and hence $I''(y)=
\partial^2_y \Phi(1,y)$. Hence, as $y\to y^*$ from above, $I''(y\to
y^*)= \partial^2_y \Phi(1,y)\Big|_{y=y^*}$. Comparing this with
\eref{h22_supp}, we see that generically the second derivative of the
rate function is discontinuous across $y=y^*$ and the value of the
discontinuity is given by
\begin{align}
\lim_{\epsilon\to 0}\,
&\left[I''(y^*-\epsilon)-I''(y^*+\epsilon)\right]= 
\Bigl[\partial_{w^*}^2 \Phi(w^*,y)\, \left[(w^*)'(y)\right]^2 \notag\\
&\qquad
+ 2\, \partial_{w^*}\partial_y\Phi(w^*,y)\, (w^*)'(y)\Bigr]\Big|_{w^*=1, y=y^*}\, .
\label{discont_supp}
\end{align}

As an example, let us consider the case of diffusion with resetting
(the first example in the main text).  In this case,
from \eref{action-diffusion}
\begin{equation}
\Phi(w^*,y)= r\, w^* + \frac{y^2}{4\,D\,w^*} \, ,
\label{diff1_supp}
\end{equation}
where $w^*(y)= y/\sqrt{4Dr} $ (considering only the positive side).
It is then straightforward to evaluate the discontinuity in
\eref{discont_supp} and we get
\begin{equation}
\lim_{\epsilon\to 0}\, \left[I''(y^*-\epsilon)-I''(y^*+\epsilon)\right]=-\frac{1}{2D}\,.
\label{diff2_supp}
\end{equation}
Similarly, one can obtain the value of the discontinuity in the second
derivative for the other examples discussed in the main text.

\end{document}